\documentclass[aps,preprint,pra,showpacs]{revtex4}
\usepackage{graphicx}

\usepackage{amsmath}
\usepackage{latexsym}
\usepackage{amsthm}

\usepackage{amssymb}

\usepackage[pdftex]{color}

\begin{document}

\title{A nonlocal ontology underlying the time-symmetric Heisenberg representation}

\author{Yakir Aharonov$^{1,2}$, Tomer Landsberger$^1$ and Eliahu Cohen$^{1,3}$}
\affiliation{$^1$School of Physics and Astronomy, Tel Aviv University, Tel Aviv 6997801, Israel\\ $^2$Schmid College of Science, Chapman University, Orange, CA 92866, USA\\ $^3$H.H. Wills Physics Laboratory, University of Bristol, Tyndall Avenue, Bristol, BS8 1TL, U.K}

\date{\today}

\begin{abstract}
We maintain that the wavefunction is an ensemble property rather than an individual particle property. For individual particles, we propose an ontology underwritten by the Heisenberg representation. It consists of properties represented by deterministic operators, which may have nonlocal dynamics. Relying on nonlocal dynamics, we show how interference phenomena can be understood without having to conceive of the quantum state as wave-like. Nonlocal information is provided by a modular momentum operator. By augmenting the individual particle ontology with a final state and employing weak measurements, we show how both interference and which-path can be deduced for the same system. This situation is most intuitively understood through a time-symmetric Heisenberg representation. Indeed, we contend that a time-symmetric operator-based ontology captures the essence of quantum mechanics, particularly its nonlocal nature, better than any wavefunction-based ontology.
\end{abstract}


\maketitle

\section{Introduction}
\label{Hintroduction}

The Schr\"{o}dinger and Heisenberg representations of quantum mechanics are mathematically equivalent, but they are very different in some important respects. While the Heisenberg formalism closely resembles its classical counterpart, namely Hamiltonian classical mechanics, the Schr\"{o}dinger formalism is regarded as simpler, admitting a physical interpretation more naturally. The central notion of the latter formalism is the {\it wavefunction}, which as its name suggests, features wave-like properties. Consequently, quantum mechanics was understood to entail that microscopic bodies have a dual wave-particle nature, construed by Bohr and others as the essence of the theory, and in fact, its main novelty. This conception is also reflected in the questions traditionally posed with respect to the foundations of quantum mechanics - is the wavefunction real? Is it a complete description of nature? Is wavefunction collapse reducible to a deterministic process? And so on \cite{Ring,Col,Gao,Wal}. Moreover, the axioms of quantum mechanics are usually stated in the Schr\"{o}dinger formalism, which is accordingly adhered to in university classroom teachings. \\

We believe the primacy of Schr\"{o}dinger representation to be the consequence of an unfortunate historical turn of events, and wish to correct that distortion. We propose an alternative ontology for quantum mechanics, which relies on the Heisenberg representation. Within it, the basic, primitive, physical properties will be represented by a set of deterministic operators, which are operators whose measurements do not disturb each other and have deterministic outcomes. The modular momentum operator will arise as particularly significant in explaining interference phenomena. It departs from its classical counterpart by having nonlocal dynamics. Indeed, dynamical nonlocality will take center stage. This sort of nonlocality, which enters the theory through the unitary equations of motion, is distinguished from the more familiar kinematical nonlocality (implicit in entangled states \cite{Bell}) in having an observable effects on probability distributions (unlike e.g. measurements of one out of two spins in a Bell states). Kinematic nonlocality was previously analyzed in the Heisenberg picture by Deutsch and Hayden \cite{DH}. Apart from a short note, this work will focus on dynamical nonlocality. In the Schr\"{o}dinger representation, this form of nonlocality is manifest in the unique role of {\it phases}, which while unobservable locally, may influence interference patterns. In addition, a final state will be considered juxtapose the usual initial state, forming to a two-fold set of deterministic properties. The above amounts to a time-symmetric Heisenberg-based ontology.\\

The article is structured as follows: Sec. I motivates an ontological alternative to the wavefunction; Sec. II introduces the requisite mathematical machinery - deterministic operators; Sec. III presents the final state as an essential ingredient in the new ontology, and its operational consequences - weak values and the extended set of deterministic operators; Sec. IV introduces the modular momentum, a dynamically nonlocal operator taking the place of relative phase within the Heisenberg representation; Sec. V explains why a two-slit-type experiment involving post-selection is better understood through the Heisenberg prism as opposed to Schr\"{o}dinger, that is, as a manifestation of nonlocal modular momentum exchange rather than of the wave nature of particles; Sec. VI is a discussion, tying together the above. The main novelty of the work is exposing the ontological underpinnings of the Heisenberg representation, and combine them with the notion of a final state into a nonlocal, deterministic, time-symmetric ontology.

\section{The wavefunction represents an ensemble property}
\label{Hthermo}

The question of the wavefunction's meaning is core in the corpus of the interpretational controversy surrounding quantum mechanics. We take neither the standard ontic nor the epistemic approach towards the wavefunction. Rather, we consider it to represent an {\it ensemble} property, as opposed to a property of an individual systems. This resonates with the {\it ensemble interpretation} of the wavefunction, initiated by Born \cite{MBorn}, and extensively developed by Ballentine \cite{Ball1,Ball2}. According to this interpretation, the wavefunction is a statistical description of a hypothetical ensemble, from which the probabilistic nature of quantum mechanics stems directly. It does not apply to individual systems. Balletine justifies an adherence to this interpretation by observing that it overcomes the measurement problem - by not pretending to describe individual systems, it avoids having to account for {\it state reduction} (collapse). We concur with Ballentine's conclusion, but do not follow his reasoning. Instead, we contend that the wavefunction is appropriate as an ensemble ontology rather than an individual system ontology, because it can only be {\it directly verified} on the ensemble level. By ``directly verified" we mean measured to an arbitrary accuracy in an arbitrarily short time (excluding practical and relativistic constraints).\\

Indeed, we only regard directly verifiable properties to be intrinsic. Consider for instance how probability distributions relate to single particles in statistical mechanics. We can measure, e.g., the Bolztmann distribution, in two ways - either instantaneously on thermodynamic systems or using prolonged measurements on a single particle coupled to a heat bath. We do not attribute the distribution to single particles because instantaneous measurements performed on single particles yield a large error. Conversely, when the system is large, containing $N>>1$ particles (the thermodynamic limit), the size of the error, which scales like $\sqrt{N}$, is relatively very small. In other words, the verification become direct as the system grows. Because of this, the distribution function is best viewed as a property of the entire thermodynamic system. On the single particle level, it manifests itself as probabilities for the particle to be found in certain states. However, the intrinsic properties of the individual particle are those which can be verified directly, namely position and momentum, and only they constitute its real properties.\\

Similarly to how distributions in statistical mechanics can only be directly verified on a thermodynamic system, the wavefunction can only be directly verified on quantum ensembles. Continuing the analogy, on a single particle level, the wavefunction can only be measured by performing a prolonged measurement. This prolonged measurement is a {\it protective} measurement \cite{prot1}. Protective measurements can be implemented in two different ways: the first is applicable for measuring discrete non-degenerate energy eigenstates and is based on the adiabatic theorem \cite{Fock}; the second, more general way, requires an external protection in the form of the quantum Zeno effect \cite{Zeno}. In either of the two ways, a large number of identical measurements is required for approximating the wavefunction of a single particle. We conclude that in analogy to statistical mechanical distributions being thermodynamic system properties, the wavefunction is a quantum ensemble property.\\

Unlike Born, we do not wish to imply that the description by means of a wavefunction is incomplete, concealing a classical reality (i.e. hidden variables); nor do we oppose the consequence of PBR theorem \cite{PBR}, which states that the wavefunction is determined uniquely by the physical state of the system. We only mean to suggest that the wavefunction cannot constitute the primitive ontology of a single quantum particle/system. That being said, contrary to ensemble interpretation advocates, we will not duck out of proposing a single-particle ontology. In what follows, we expound such an ontology based on {\it deterministic operators}, which are unique operators whose measurement can be carried on a single particle without disturbing it, with fully predictable outcomes. Since properties corresponding to these operators can be directly verified on the single particle level, they constitute the {\it real} properties of the particle. In order to derive this ontology, we turn the spotlight to the Heisenberg representation.

\section{Deterministic operators represent single-particle properties}
\label{Hdet}

In the Schr\"{o}dinger picture, a physical system is fully described by a ray in a Hilbert space, or, in the position representation, by a continuous wavefunction. Its evolution is dictated by the Hamiltonian and calculated according to Schr\"{o}dinger equation. The observables are described by operators, which remain idle in this representation. In the Heisenberg representation, a physical system can be described by a closed (under addition and multiplication) set of {\it deterministic operators}, evolving according to Heisenberg equation, whereas the wavefunction remains idle. Deterministic operators are Hermitian ``eigenoperators", that is, Hermitian operators for which the system’s state is an eigenstate:

\begin{equation}
\{A_i ~\text{such that}~ A_i|\psi\rangle=a_i|\psi\rangle, a_i \in \mathbb{R}\}.
\end{equation}

To describe a particle in an $n$-dimensional Hilbert space, a set of $(n-1)^2+1$ deterministic operators, whose eigenvectors span the relevant sub-space, is required \cite{Toll2009}. The physical significance of these operators stems from the possibility to measure them without disturbing the particle, i.e. without inducing collapse. That being the case, they can also be measured successively without mutual disturbance

\begin{equation}
[A_i, A_j]|\psi\rangle=0,
\end{equation}

\noindent for any $i,j$. \\

As long as only eigenoperators are measured, they evolve unitarily according to the Heisenberg equation applied separately to each of them. However, when a projective measurement of an operator which does not belong to this set is performed, the set of deterministic operators is redefined.\\

Deterministic operators, whose measurement outcomes are completely certain, are dual to {\it completely uncertain operators}, whose measurement outcomes, as the name suggests, are completely {\it un}certain. This means that they satisfy the condition that all their possible measurement outcomes are equiprobable \cite{AR}. Thus no information can be gained by measuring them. The importance of this feature will become clearer later on.\\

The mathematical equivalence between the Schr\"{o}dinger and Heisenberg pictures assures that there is a one-to-one correspondence between the wavefunction and the set of deterministic operators describing the same physical system. But the wavefunction also expresses non-deterministic properties, such as positions in a delocalized system. We do not consider these to be real properties of the single particle, maintaining that they are exhausted by the set of deterministic properties. If, for example, the position operator is not deterministic, the question ``where is the particle?'' bares no meaning in that context. We will see that in all cases where the position of a particle is actually measured, it will be represented by a deterministic operator. The non-deterministic operators do not represent properties intrinsic to the particle, only to an ensemble of similarly prepared particles. Indeed, an ensemble possess a set of deterministic operators larger than that of the single particle. The average value of any one-particle operator is deterministic, for instance, as a result of the law of large numbers. Importantly, the wavefunction itself is a deterministic operator of an ensemble of such particles according to the same logic. \\

\section{The final state}
\label{Htsvf}

An important ingredient of the proposed ontology is a final state describing the system together with the usual initial state. The idea that a complete description of a quantum system at a given time must take into account two boundary conditions rather than one is known from the two-state vector formalism (TSVF). The TSVF is a time-symmetric formulation of standard quantum mechanics, which posits, in addition to the usual state vector, a second state vector evolving from the future towards the past. This approach has its roots in the works of Aharonov, Bergman and Lebowitz \cite{ABL}, but it has since been extensively developed \cite{TSVF}, and has led to the discovery of numerous peculiar phenomena \cite{AR}. \\

The TSVF provides an extremely useful platform for analyzing experiments involving pre- and post-selected ensembles. Post-selection is permitted in quantum mechanics due to the effective indeterminacy of measurement, which entails that the state of a system at one time and its Hamiltonian only partially determine measurement outcomes at later times. {\it Weak measurements} enable us to explore the state of the system at intermediate times without disturbing it \cite{AAV,4f}. This type of measurement is based on a very weak von Neumann coupling to a pointer, such that the measured state is negligibly disturbed. The two-state $\langle \phi |~| \psi \rangle$ created by both boundary conditions allows to define for any operator the {\it weak value} by:

\begin{equation}
\langle A \rangle_w= \frac{\langle \phi |A| \psi \rangle}{\langle \phi | \psi \rangle}.
\end{equation}\\

A single weak measurement provides a negligible amount of information, but when repeated over a large pre- and post-selected ensemble, weak measurements can reveal the weak value with high accuracy \cite{ACE,4f}. The power to explore the pre- and post-selected system by employing weak measurements motivates a literal reading of the formalism, that is, as more than just a mathematical tool of analysis. It motivates a view according to which future and past are equally important in determining the quantum state at intermediate times, and hence equally {\it real}. Accordingly, in order to fully specify a system, one should not only pre-select, but also post-select a certain state using a projective measurement.\\

We wish to consider a final state in the framework of the Heisenberg representation. In that formalism, adding a final state amounts to adding a second set of deterministic operators on top of the one dictated by the initial state, thereby enlarging the assortment of system properties. The properties expressed by this two-fold set are the ones which we regard to constitute the primitive ontology of quantum mechanics.

\section{Wave-like behavior, relative phase and modular operators}
\label{HMod}

Interference patterns appear in both classical and quantum grating experiments (most conveniently analyzed in a two-slit setup, which will be referred to hereinafter). Allegedly, an explanation to interference phenomena is shared across the domains: a spatial wave(function) traverses the grating, one part of whom goes through the first slit while the other goes through the second, before the two parts meet to create the familiar pattern. Indeed, it is tempting to extend the accepted classical explanation into the quantum domain. But before jumping on this bandwagon, an important dis-analogy should be noted. In a classical wave theory, one can deduce what will happen when the two parts of the wave finally meet based on {\it local} information available along the trajectories of the wavepackets going through the two slits. In quantum mechanics however, what tells us where the maxima and minima of the interference are, is the {\it relative} phase of the two wavepackets. Crucially, the two local phases cannot be observed, for a measurement revealing the local phase would violate gauge symmetry \cite{AR}. Only the phase difference is observable, but it cannot be deduced from measurements performed on the individual wavepackets. The analogy is partial. For this reason, we contend that the temptation to jump on the wavefunction bandwagon should be resisted.\\

We would like to show how interference can be understood without having to say that each particle passed through both slits as if it were a wave. For this purpose, we examine what are the operators sensitive to the property which determines the interference pattern in Schr\"{o}dinger picture, namely the relative phase. Interestingly, the position and momentum operators, and every finite polynomial of them, are not \cite{AR}. If we define the freely evolving $\Psi_\phi(x,0)=\psi_1(x,0)+e^{i\phi}\psi_2(x,0)$, where $\psi_1(x,0)$ and $\psi_2(x,0)$ are non-overlapping wavepackets a distance $L$ apart, then

\begin{equation}
\int_{-\infty}^{\infty} \Psi^*_\phi(x,t)x^mp^n\Psi_\phi(x,t)dx
\end{equation}\\

\noindent is independent of $\phi$ for every $t$ and for all $m$ and $n$. This is simply because $\psi_1(x,0)$ and $\psi_2(x,0)$ remain non-overlapping after any application of $x$ and $p$. It turns out however, that this relative phase can be measured through the {\it modular momentum} operator $p_{mod}$ \cite{APP}. In a two-slit model with separation $L$ between slits, it is defined to be $p$ mod $\hbar/L$, which takes on values in the interval $[0, \hbar/L)$. For deductive purposes, we employ the equivalent $e^{ipL/\hbar}$ which clearly depends only on $p_{mod}$. The expectation value of this operator reveals the relative phase between the two wavepackets

\begin{equation}
\int_{-\infty}^{\infty} \Psi^*_\phi(x,t)e^{ipL/\hbar}\Psi_\phi(x,t)dx=e^{i\phi}/2.
\end{equation}\\

It is generally true that the operators sensitive to the relative phase are periodic functions of $x$ and $p$, and not merely polynomials. The apparent paradox posed by Taylor theorem is dissolved by observing that the expectation value of an infinite sum does not equal the sum of expectation values. Those operators can be expressed as functions of the modular momentum operator. An interesting feature of the modular momentum operator is that it evolves according to {\it nonlocal equations of motion}. For instance, when $H=\frac{p^2}{2m}+V(x)$ the time evolution is

\begin{equation}
\frac{d}{dt}e^{ipL/\hbar}=\frac{i}{\hbar}[V(x)-V(x+L)]e^{ipL/\hbar},
\end{equation}\\

\noindent which depends on the two, possibly remote positions $x$ and $x+L$. This departs considerably from the classical evolution according to Poisson brackets

\begin{equation}
\frac{d}{dt}e^{i 2\pi p/p_0}=\left \{e^{i 2\pi p/p_0},H\right\}=-i\frac{2\pi}{p_0}\frac{dV}{dx}e^{i 2\pi p/p_0},
\end{equation}\\

\noindent which involves a {\it local derivative} (we defined $p_0\equiv h/L$), suggesting that the classical modular momentum changes only if a local force $\frac{dV}{dx}$ is applied. We thus understand that commutators, although having a classical limit in terms of Poisson brackets, are inherently different, entailing nonlocal dynamics. This is to be expected since the modular momentum operator lacks a classical analog, diverging in the limit of $\hbar \rightarrow 0$. The connection between nonlocal dynamics and relative phase via the modular momentum clues to the possibility of the former taking the place of the latter in the Heisenberg representation. We will show that this is indeed the case.\\

As defined by Aharonov and Rohrlich \cite{AR}, the {\it complete uncertainty principle} states that any periodic function $\Phi$, and specifically the modular momentum which is periodic in $p$, is completely uncertain (that is, its values are equiprobable) if and only if $\langle e^{in\Phi}\rangle$ vanishes for every integer $n$. Taking the Fourier expansion of $Prob(\Phi)=\sum_{n=-\infty}^{n=+\infty}a_n e^{in\Phi}$, where $a_n=\int Prob(\Phi) e^{in\Phi}d\Phi=\langle e^{in\Phi}\rangle$, we see that $Prob(\Phi)=const.$ if and only if $\langle e^{in\Phi}\rangle=0$ for all $n \neq 0$.\\

When a particle is localized to within $|x|<L/2$, the expectation value of $e^{ipL/\hbar}$ vanishes. This is obvious since $e^{ipL/\hbar}$ functions as a translation operator, shifting the wavepacket outside $|x|<L/2$. Accordingly, when a particle is localized near one of the slits, as in the case of $\psi_1$ and $\psi_2$, $\langle e^{inpL/\hbar}\rangle=0$ for every $n$. It then follows from the complete uncertainty principle that its modular momentum is completely uncertain. Accordingly, all information about the modular momentum is lost once we find the position of the particle. The unset of complete uncertainty is crucial in order to prevent signaling and preserve causality. To realize this, consider that if we apply a force arbitrarily far away from a localized wavepacket, we can change its modular momentum instantly, since modular momentum relates remote points in space. Measuring this change on the wavepacket could then allow a violation of causality. Fortunately, this measurement is precluded by complete uncertainty.\\

The fact that the modular momentum becomes uncertain upon localization accords well with the fact that interference is lost in that case. In the Schr\"{o}dinger picture, interference loss is understood as a consequence of wavefunction collapse. Once the superposition is reduced, there is nothing left for the remaining wavepacket to interfere with. In the Heisenberg picture, collapse cannot be described. But the above suggests a Heisenbergean physical explanation for interference loss. If one of the slits is closed by the experimenter, a nonlocal exchange of modular momentum with the particle occurs. Consequently, the modular momentum becomes completely uncertain, thereby erasing interference vis a vis destroying the information about the relative phase.\\

Since $p= p_{mod}\ +N\hbar/L$ for some integer $N$, the uncertainty of $p$ is greater or equal to that of $p_{mod}$ (the integer part can be uncertain as well). For this reason, a complete uncertainty of the modular momentum $p_{mod}$ (which means its distribution function is uniform in the interval $[0,\hbar/L)$) sets $\hbar/L $ as a lower bound for the uncertainty in $p$, i.e. $\Delta p \ge \hbar/L$. This inequality parallels the Heisenberg uncertainty, equating it in the case of $\Delta x = L$, which is why we regard the complete uncertainty principle as a very fundamental one.

\section{Pre- and post-selected interference experiments comply better with Heisenberg}
\label{Imply}

Performing certain experiments involving post-selection will allow us both to measure interference and deduce which-path information. Such experiments are not well understood through the Schr\"{o}dinger picture, since it becomes necessary to posit wave and particle properties {\it at the same time}. Alternatively, in the Heisenberg picture, the particle has a determinate location, alongside a nonlocal modular momentum, which can ``sense'' the presence of the other slit and bring about interference. This description escapes the difficulty present in the Schr\"{o}dinger picture. \\

\begin{figure}[b!]
\includegraphics[height=6.8cm]{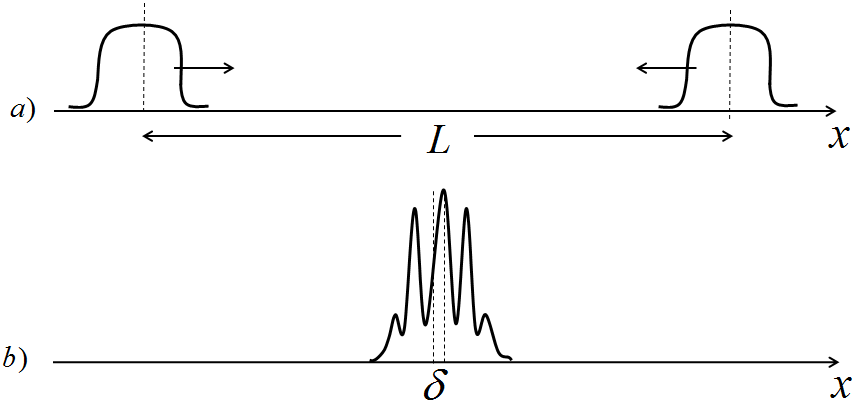}\\
\caption{{\bf Interference of two wavepackets.} a) The density of the initial superposition (\ref{state}) of the two wavepackets. b) The interference pattern at the time of the exact overlap between envelops of the wavepackets. The shift $\delta$ of the interference pattern is proportional to the relative phase $\phi$. } \label{1}
\end{figure}

Let us consider simple Gedanken experiment, which will first be described in the Schr\"{o}dinger representation. A particle is prepared in a superposition of two identical spatially separated wavepackets moving toward one another with equal velocity (Fig. \ref{1}). This can be written as

\begin{equation}\label{state}
\Psi(x) = \frac{1}{\sqrt{2}}[e^{ip_0x/\hbar}\Psi_1(x + L/2)+ \Psi_2(x)],~~~{\rm } ~~\Psi_2(x) = e^{i\phi}e^{-ip_0x/ \hbar}\Psi_1(x - L/2),
\end{equation}\\

\noindent where $\Psi_1$ is some localized (and normalized) wavefunction. The relative phase $\phi$ has no effect on the local density $\rho(x)$ or any other local feature until the two wavepackets overlap. The phase manifests itself by shifting the interference pattern by $\delta=\frac{\hbar\phi}{p_0}$.\\

This initial configuration is similar to that of the two-slit setup, but instead of letting the two wavepackets propagate away from the grating to hit a photographic plate, we let them meet at time $T$ on the plane of the grating. Upon meeting, the density of the two wavepackets becomes

\begin{equation}
\rho(x,T)=4|\Psi_1(x)|^2\text{cos}^2(p_0x/\hbar+\phi/2),
\end{equation}\\

\noindent which spells interference, similar to that of a standard two-slit experiment.\\

We now augment the experiment with a post-selection procedure, where we place a detector on the path of the wavepacket $\Psi_2$, moving to the right. The probability of finding the particle there is $\frac{1}{2}$. Let us consider an ensemble of such pre- and post-selection experiments, which realizes the rare case where {\em all} the particles are found by this detector (that is, we determine the position operator for the entire ensemble by a post-selection). The two-state, which constitutes the full description of pre- and post-selected systems at any intermediate time $t$, is given by $\langle \Psi_2(t)|~|\Psi(t)\rangle$. Within the TSVF, we can define a two-times generalization of the pure-state density:

\begin{equation}
\rho_{two-time}(x,T)=\frac{\langle x|\Psi\rangle\langle \Psi_2|x\rangle}{\langle \Psi | \Psi_2 \rangle}=2|\Psi_1(x)|^{2}e^{i(p_0x/ \hbar-\phi/2)}\text{cos}(p_0x/\hbar-\phi/2).
\end{equation}\\

\begin{figure}[t!]
\centering\includegraphics[height=2.6cm]{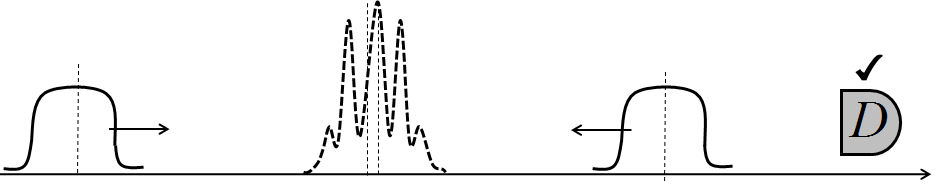}\\
\caption{{\bf Weak measurement of the interference pattern.} Weak measurements show the usual interference pattern in spite of the fact that detector $D$ detects all particles as belonging to just one (moving to the right) wavepacket. } \label{3}
\end{figure}

To measure this density, we perform (at intermediate times) a weak measurement using $M>>1$ projections $\Pi_i(x)$ with the interaction Hamiltonian $H_{int}=g(t)q\sum_i^M\Pi_i(x)$, where $q$ is the pointer of the measuring device, $i$ sums over an ensemble of particles, and $\int_0^\tau g(t)dt=g$ is sufficiently small during the measurement duration $\tau$. For a large enough ensemble, these measurements allow us to observe the two-time density while introducing almost no disturbance to the modular momentum of the particles. If we perform many such measurements in different locations within the overlap region, they will add up to a histogram tracing the two-time density in that region (Fig. \ref{3}). From this histogram we can find the parameter $\delta$ which depends on the relative phase $\phi$. This experiment demonstrates a rather perplexing situation - The real part of this density, which determines the weak measurement outcomes, is identical to that of the regular density of the two wavepackets. Therefore, weak measurements of projections performed on this system will also exhibit an interference pattern. However, by virtue of the post-selection, we know that the particle is described by a right-moving wavepacket which went through the left slit. Thus we observe the two-slits-like interference pattern, but nevertheless know through which slit the particles have passed. Indeed, analyzing this experiment through the prism of the Schr\"{o}dinger representation is rather confusing. How can each particle have a well-defined position, and also exhibit an interference pattern? \\

In contrast, the Heisenberg representation tells us that each particle had a determinate position, but at the same time also held nonlocal information in the form of a deterministic operator depending on the modular momentum. In the special case of a two-slit interference experiment, the deterministic operators sensitive to the relative phase can be constructed out of the following basis set:

\begin{equation}
\begin{array}{lcl}
\sigma_1=\text{cos}(pL/\hbar)-\text{sin}(pL/\hbar)\frac{\text{sin}(\pi x/L)}{|sin(\pi x /l)|} \\
\sigma_2=\text{sin}(pL/\hbar)+i\text{cos}(pL/\hbar)\frac{\text{sin}(\pi x/L)}{|sin(\pi x /l)|}.\\
\end{array}
\end{equation}

Together with\\

\begin{equation}
\sigma_3=\frac{\text{sin}(\pi x/L)}{|\text{sin}(\pi x /l)|},
\end{equation}

\noindent which reveals the position (a local property), the trio of operators form a set of spin-half-like observables \cite{ Toll2009}:\\

\begin{equation}
[\sigma_i, \sigma_j]=2i\epsilon_{ijk}\sigma_k.
\end{equation}

Specifically, the deterministic operator for the case of a relative phase $\alpha$ is

 \begin{equation}
\sigma_1\text{cos}(\alpha)+\sigma_2\text{sin}(\alpha)=1.
\end{equation}

The emerging picture might be harder to visualize (and by consequence - sketch), but by exercising this thinking, we believe one may gain a new and powerful intuition about the underlying physics.\\

\section{Discussion}
\label{Hconc}

The Heisenberg-based ontology has some clear advantages over a Schr\"{o}dinger-based ontology. In the Heisenberg picture, the point of departure from classical mechanics is clearly visible - the transition from Poisson brackets to Moyal brackets (commutators) introduced nonlocal equations of motion. Nonlocality, which will later emerge even more dramatically in the EPR experiment, is presented by properties that have no classical analog. These properties can be associated with each individual particle, unlike the Schr\"{o}dinger wave, which is a property of an ensemble. They enable us to understand interference in experiments involving both pre- and post-selection, where a particle exhibiting interference also has a definite position.\\

In this ontology, an uncertainty principle appears not as a mathematical consequence, but as a reconciler between a metaphysical desiderata - causality, and the nonlocality of the dynamics. This complete uncertainty principle (qualitatively) implies the Heisenberg uncertainty principle, but not the other way around. For these reasons, we regard as more fundamental. In turn, uncertainty combined with a demand for single-valued measurement outcomes, necessitates a mechanism for choosing those outcomes. We have shown elsewhere \cite{Collapse}, that by considering a special final state of the kind we had introduced, but for the entire Universe, the outcomes of specific measurements can be accounted for. This cosmological generalization thereby answers the {\it measurement problem}. We now understand this final state to constitute a set of deterministic properties, which may be regarded as a hidden variables due to its epistemic inaccessibility in earlier times.\\

By defining a system using both pre- and post-selection, a broader notion of a physical state is obtained, dictating not just the expectation values of all operators, but also which properties belong to the particle (deterministic) and which do not (non-deterministic). The intrinsic nature (that is, set of properties) of the particle therefore depends on both preparation and post-selection, unlike the case of classical mechanics. The real properties of the particle consist of a conjunction of two sets of deterministic properties, which may be nonlocal. These properties evolve deterministically in accordance with Heisenberg equation. This interpretation of quantum mechanics is thus both deterministic (in a broad two-times sense) and nonlocal. Average values of operators become deterministic and therefore intrinsic on the ensemble level. In this way, the macroscopic world emerges from the microscopic one.\\

Internalizing this ontology, one is no longer restricted to thinking in terms of the Schr\"{o}dinger representation, which is very convenient mathematically but often confusing intuitively. Given, the wavefunction is an efficient mathematical tool for calculations of experiment statistics. But potential functions too are mathematically efficient, whereas it is only the field derived from them which is physically real. Hence mathematical usefulness is not a sufficient condition for reality. Indeed, while useful for calculating the dynamics of the deterministic operators, the wavefunction is not the real physical object - only the deterministic operators themselves are. Importantly, considerations pertaining to this ontology have led Aharonov to discover the Aharonov-Bohm effect. The stimulation of new discoveries is the ultimate trial of an interpretation.\\

Intriguingly, the Heisenberg representation which was discussed here from a foundational point of view, is also a very helpful framework for discussing quantum computation \cite{Gott1998}. Moreover, in several cases \cite{Gard2014}, it has a computational advantage over the Schr\"{o}dinger representation.\\

For the sake of completeness, it might be interesting to briefly address the notion of kinematical nonlocality arising from entanglement. As noted in Sec. III, a quantum system in two-dimensional Hilbert space, e.g. a spin-1/2 particle, is described within our formalism using two deterministic operators. For describing a system of two entangled spin-1/2 particles (in a four-dimensional Hilbert space), we would utilize a set of 10 deterministic operators. It is important to note that the measurements of such operators are nonlocal \cite{Vaidnonlocal}, possibly carried out in space-like separated points. Most of these operators involve simultaneous measurements of the two particles. A (non-deterministic) measurement of one particle would change the combined set of deterministic operators, thus instantaneously affecting also the ontological description of the second particle. For another, information-based perspective on this subject, we refer the reader to \cite{DH}. There it was claimed that the information flow in the Heisenberg representation is local, however, in light of the above analysis, this only refers to certain kinds of operators.\\

A note of conclusion - We believe that had quantum mechanics preceded relativity theory, the proposed ontology could have been the commonplace one. Before the 20th century, physicists and mathematicians were interested in studying various Hamiltonians having an arbitrary dependence on the momentum, such as $\text{cos}(p)$. In quantum mechanics, these Hamiltonians lead to nonlocal effects as discussed above. The probability current is not continuous under the resulting time-evolution, which makes the wavefunction description less intuitive. However, those Hamiltonians were dismissed as non-physical in the wake of relativity theory, allowing the wavefunction ontology to prosper. We hope that our endorsement of the Heisenberg-based ontology will promote a discussion of this somewhat neglected approach.\\

\break

\begin{acknowledgments}

We cordially thank Daniel Rohrlich for many helpful discussions. Y.A. thanks the Israel Science Foundation (grant no. 1311/14), the ICORE Excellence Center ``Circle of Light'' and the German-Israeli Project Cooperation (DIP) for support. E.C. was supported by ERC AdG NLST.

\end{acknowledgments}

\end{document}